\documentstyle[preprint,aps, epsfig]{revtex}
\begin{document}
\input{psfig.sty}
\def \gam {\frac{ N_f N_cg^2_{\pi q\bar q}}{8\pi} }
\def \gamm {N_f N_cg^2_{\pi q\bar q}/(8\pi) }
\def \be {\begin{equation}}
\def \ba {\begin{eqnarray}}
\def \ee {\end{equation}}
\def \ea {\end{eqnarray}}
\def \gap {{\rm gap}}
\def \gapp {{\rm \overline{gap}}}
\def \gappp {{\rm \overline{\overline{gap}}}}
\def \im {{\rm Im}}
\def \re {{\rm Re}}
\def \Tr {{\rm Tr}}
\def \P {$0^{-+}$}
\def \S {$0^{++}$}
\def \uu {$u\bar u$}
\def \dd {$d\bar d$}
\def \ss {$s\bar s$}
\def \qq {$q\bar q$}
\def \qqq {$qqq$}
\def \si {$\sigma(600)$}
\def \lsm {L$\sigma $M}
\title{
Footprints of a Broad $\sigma$(600) in Weak-Interaction Processes}
\author{A.D. Polosa and N. A. T\"ornqvist}\address{Physics Department,
POB 9, FIN--00014, University of Helsinki, Finland}
\author{M. D.  Scadron}
\address{Physics Department, University of Arizona, Tucson, AZ 85721,  USA}
\author{V. Elias}\address{Department of Applied Mathematics,
The University of Western Ontario, London, ON  N6A 5B7, Canada}
\maketitle
\begin{abstract}
We explore how chiral-symmetry constraints on
weak-interaction matrix elements point toward
the existence of an intermediate-state $\sigma$ in
several different weak-interaction processes.
Particular attention is directed toward recent
evidence for a $\sigma$ within three-body
nonleptonic weak decays.\\

\noindent PACS numbers: 11.30.Rd, 11.40.Ha, 14.40.Cs\\
\noindent HIP-2000-22/TH
\vskip 0.90cm
\end{abstract}

The existence of a broad scalar-isoscalar
$\pi \pi$ resonance near 600 MeV has long been
controversial.  After having been absent for many years
from the listings of the Particle Data Group (PDG) in the Reviews of
Particle Physics\cite{pdg}, this resonance has reappeared under the
name ``$f_0(400-1200)$ or $\sigma$" (hereafter called \si ) after
reanalysis by several groups\cite{due} of the available data on the
scalar nonet and $\pi\pi$ phase shifts. Today an increasing number
of theoretical and experimental analyses point toward the existence of
this important meson\cite{tre,quattro}.  A salient feature of such
analyses is chiral symmetry and its soft breaking.
Essentially all groups who have recently analyzed the strong-interaction
scalar meson data within a chiral framework (together with
unitarity, analyticity, flavour symmetry etc.) seem to require the
existence of the \si.   The simplest model 
in which $\sigma$ occurs is the linear sigma model ($L \sigma M$),
a model which
implements chiral symmetry for scalar and pseudoscalar mesons (and
unconfined quarks) together with flavour symmetry. Below we shall appeal to several
applications of this model to weak decays, as naive quark models
without chiral-symmetry constraints are unsuccessful in describing properties of the
lightest scalars.  Similarly, the nonlinear sigma model (and chiral-perturbation-theory approaches
which follow from it) can be understood as an $m_\sigma \rightarrow
\infty$ $\sigma$-model limit, a limit appropriate at very low energies
but inappropriate for processes whose momenta are comparable to the
$\sigma$-mass. \footnote{For example, Sannino and Schechter [5] utilise
chiral-perturbation-theory to predict the initial behaviour of the
$\pi\pi$-scattering amplitude for the sub-400 MeV region, but in
subsequent work with Harada [4] see clear evidence for a 
$\sigma$-resonance past that region.}

The large width of the \si\ (over $300\;{\rm MeV}$) has led many
to argue that its nature as a resonance is obscured, and that
the $\pi\pi$ phase shift in the $500-900\;{\rm MeV}$ region
may arise solely from the large contributions of crossed channel
diagrams\cite{sei}. However, the duality
between {\it s-} and {\it t-}channel exchanges
(Regge poles), which has been well-established for more than two decades,
indicates that strong crossed-channel effects always appear
in conjunction with a resonance.

In this paper we shall briefly review the evidence for  the \si\ in weak
interaction decays. Although this evidence, like
that from strong interactions, is largely indirect and contingent
upon the theoretical framework for incorporating chiral symmetry,
we nevertheless find that analyses of several different
processes point toward the existence of the {\si}, 
as predicted by {\lsm} physics.  Thus direct
experimental proof of the $\sigma$'s existence is becoming more
and more pressing. The recent data analysis\cite{sette} on
$D\to\sigma\pi\to 3\pi$ from the E791 experiment at Fermilab, as
discussed in the final section of this paper, appears to provide more 
direct weak-interaction evidence for a physical $\sigma$ resonance.

\newcommand{\e}{{{e}}}
\begin{center}{\it{\lsm} and ${(\pi^+, K^+)\to \e^+\nu \gamma}$ Semileptonic Weak Decays}
\end{center}
Let us first review some {\lsm} evidence for a $\sigma$
within the $\pi^+\to e^+\nu \gamma$ decay. The $uud$ quark
triangle graph of Fig. 1(a) is known \cite{otto} to predict a value of unity
for the structure-dependent axial-to-vector form factor ratio at
$q^2=0$:
\be
\gamma_q =\frac{F^q_A(0)}{F^q_V(0)}=1. \label{eq:uno} \ee
Within a \lsm\ framework for chiral symmetry breaking, one also must
consider triangle graphs involving mesons \cite{nove}, as in Fig. 1(b), in which case the
net $q^2=0$ form factor ratio is reduced to 
\be
\gamma_{\rm L\sigma M} =\frac{F^{\rm L\sigma M}_A(0)}{F^{\rm L\sigma
M}_V(0)}=1-\frac 1 3 =\frac{2}{3}. \label{eq:due}
\ee
The $q^2=0$ limit of the axial-to-vector form factor
ratio has also been extracted from experimental data \cite{pdg}:
\be
\gamma_{\rm PDG} =\frac{F_A(0)}{F_V(0)}=\frac{0.0116\pm
0.0016}{0.017\pm 0.008}=0.68\pm 0.33. \label{eq:tre} \ee The
agreement between (\ref{eq:due}) and the central value of
(\ref{eq:tre}) suggests the inclusion of $\pi$ and $\sigma$ within
the low-energy effective theory, although the empirical range
(\ref{eq:tre}) cannot be said to exclude the meson-free prediction
(\ref{eq:uno}).

In a similar manner, at $q^2=0$ the $SU(3)_f$ \lsm\ applied to
$K^+\to e^+\nu \gamma$  has been shown to predict \cite{nove}
that
\be
|F^K_V(0)+F^K_A(0)| \approx (0.22+0.09)\;{\rm GeV}^{-1}=0.31\;{\rm
GeV}^{-1}\ . \label{eq:quattro} \ee Once again the \lsm\ meson
loops reduce $F_A$ with respect to $F_V$ because of the relative
minus sign between quark and meson loops. The result
(\ref{eq:quattro}) is obtained using {\lsm} tree-level scalar
masses $m_\sigma = 680 \;{\rm MeV}$, $m_\kappa = 850 \; {\rm MeV}$
away from values characterizing the chiral limit. In any case, the
present PDG\cite{pdg} average value for the $q^2=0$ $K^+\to
e^+\nu\gamma$ form-factor sum is
\be
|F^K_V(0)+F^K_A(0)|= (0.148\pm 0.010)m^{-1}_K = (0.30\pm
0.02)\;{\rm GeV}^{-1}, \label{eq:cinque} \ee in excellent
agreement with (4).

\begin{center}{\it$\sigma$-Sensitive Matrix Elements and $K_S \to \pi\pi$ Decays.}
\end{center}

Since it is well known\cite{pdg} that $K_S\to(\pi^+\pi^-,
\pi^0\pi^0)$ decay amplitudes are overwhelmingly ($95\%$) in the $\Delta
I=\frac{1}{2}$ channel, we shall focus entirely on the $\Delta
I=\frac{1}{2}$ contribution anticipated from intermediate $\sigma$-state effects. 
>From the perspective of the {\it s-}channel, the scalar $I=0$ $\sigma (600)$  
meson (nonperturbative) pole of
Fig. 2(a) supports the $\Delta I=\frac{1}{2}$ rule \cite{dieci}
because the matrix element $\langle \sigma|H_w|K_S\rangle$ probes only the
$\Delta I=\frac{1}{2}$ component of the parity violating  weak Hamiltonian density
$H_w^{\rm pv}$. Invoking the \lsm\ vertex
\cite{undici} $\langle 2\pi|\sigma\rangle=\frac{m_\sigma^2}{2
f_\pi}$ and assuming that 
$\Gamma(\sigma)\simeq m_\sigma$ \cite{dodici}, one can estimate the $K_S\to 2\pi^0$ 
amplitude of Fig. 2(a) to be \cite{tredici,quattordici}
\begin{equation}
|\langle 2\pi^0|H_w^{\rm pv}|K_S\rangle|=\left|2\langle
2\pi|\sigma\rangle
\frac{1}{m_K^2-m_\sigma^2+im_\sigma\Gamma_\sigma}\langle
\sigma|H_w^{\rm pv}|K_S\rangle \right|\simeq\left|\frac{\langle
\sigma |H_w^{\rm pv}|K_S\rangle}{f_\pi}\right|, \label{eq:sei}
\end{equation}
($f_\pi\simeq 93$ MeV) because
$m_K^2-m_\sigma^2<<m_\sigma \Gamma_\sigma\approx m_\sigma^2$
\cite{dieci}. The value of the matrix element on the left hand side of 
(\ref{eq:sei}) can be extracted from data \cite{pdg},
\begin{equation}
|\langle 2\pi^0|H_w^{\rm
pv}|K_S\rangle|=m_K\sqrt{\frac{16\pi\Gamma_{K_S\to 2\pi}}{q}}=
(37.1 \pm 0.2)\times 10^{-8}\;{\rm GeV}, \label{eq:sette}
\end{equation}
leading via (\ref{eq:sei}) to the following estimate of the $\Delta I=\frac{1}{2}$
transition amplitude: 
\begin{equation}
|\langle \sigma|H_w^{\rm pv}|K_S\rangle|\simeq 3.45\times
10^{-8}\;{\rm GeV}^2. \label{eq:otto}
\end{equation}
The larger $K_S\to \pi^+
\pi^-$ amplitude $|\langle \pi^+ \pi^- | H_w^{\rm pv} | K_s \rangle | =
39.10 \pm 0.01 \times 10^{-8}$ GeV, analogous to 
(\ref{eq:sette}), yields a slightly larger estimate.

A corresponding estimate of the matrix element 
$< \pi | H_w^{pc} | K_L >$ can be extracted from the 
$K_S\to \pi\pi$ process in the dual {\it
t-}channel via the $K_S$ tadpole graph depicted in Fig. 2(b). The
chiral relation $[Q+Q_5,H_w]=0$ for an $H_w$ built up from $V-A$
currents generates a PCAC-consistent
$K_{2\pi^0}$ amplitude in the chiral limit
\cite{quindici,sedici,diciassette}:
\begin{equation}
|\langle 2\pi^0|H_w^{\rm
pv}|K_S\rangle|\approx
|\langle 0 |H_w^{\rm pv}|K_S\rangle
\langle K_S 2\pi^0| K_S \rangle| \frac{1}{m_{K_S}^2}
=\frac{1}{f_\pi}|\langle
\pi^0|H_w^{\rm pc}|K_L\rangle|, \label{eq:nove}
\end{equation}
where the required tadpole-PCAC transition
$|\langle {0}|H_w^{\rm pv}|K_S\rangle|=|2 f_\pi\langle\pi^0|H_w^{\rm
pc}|K_L\rangle|$ appears in the middle term of (9), and where
$\langle K_S 2\pi^0 | K_S \rangle = m_{K_S}^2 / 2 f_\pi^2$ is a matrix-element estimate
from chiral strong-interaction physics \cite{diciotto}. The right
hand sides of (\ref{eq:sei}) and (\ref{eq:nove}) thus appear to be chirally related:
\begin{equation}
|\langle\sigma|H_w^{\rm pv}|K_S\rangle| \cong |\langle \pi^0|H_w^{\rm
pc}|K_L\rangle|. \label{eq:dieci}
\end{equation}
Note that this chiral-symmetry relation (\ref{eq:dieci}) follows either from a broad
light $\sigma$ or from a narrow heavier $\sigma$;  both the former 
$\left[ \Gamma_\sigma \simeq m_\sigma, m_K^2 - m_\sigma^2 << m_\sigma^2 \right]$
and the latter  $m_\sigma^2 >> \{m_K^2, m_\sigma \Gamma_\sigma\}$ case lead to the final 
result of (6).  Moreover, if one regards equality within (10) as a fundamental constraint from chiral 
symmetry, one can reason backward from (10) to require via (6) that
$\sqrt{(m_K^2 - m_\sigma^2)^2 + m_\sigma^2 \Gamma_\sigma^2} \simeq
m_\sigma^2$, in which case
\begin{equation}
\Gamma_\sigma \simeq \frac{m_K}{m_\sigma} \sqrt{2m_\sigma^2 - m_K^2}.
\label{eq:undici}
\end{equation}
The above relation suggests that $\sigma$ is broad even if its mass is
at the high end of the 400-1200 MeV empirical range \cite{pdg} -- we see
from (11) that $\Gamma_\sigma \simeq 660$ MeV when $m_\sigma = 1$ GeV.
Note also that $\Gamma_\sigma = 570$ MeV when $m_\sigma=600$
MeV, consistent with estimates suggested in \cite{dodici}.

\begin{center}{\it $\sigma$-Sensitive Matrix Elements and Radiative Neutral Kaon Decays}
\end{center}

For the weak radiative kaon decay $K_L \to
2\gamma$, the dominant $\pi^0$ pole graph of Fig. 3(a) generates
the amplitude
\begin{equation}
\langle 2\gamma|H_w^{\rm pc}|K_L\rangle=\langle
2\gamma|\pi^0\rangle \frac{1}{(m_{K_L}^2-m_{\pi^0}^2)}\langle
\pi^0|H_w^{\rm
pc}|K_L\rangle=F_{K_L\gamma\gamma}\epsilon^{\prime\mu}\epsilon^{\nu}
\epsilon_{\mu\nu\alpha\beta}k^{\prime \alpha}k^{\beta}.
\label{eq:dodici}
\end{equation}
The near-cancellation of possible additional $\eta^{\prime}, \eta \to 2\gamma$ pole terms 
\cite{sedici,nineteen} is discussed in the Appendix to this paper.

Factoring out the Levi-Civita covariant from the $\pi^0\to
2\gamma$ amplitude, \footnote{One obtains the $\pi^0\gamma\gamma$
amplitude $\alpha/\pi f_\pi=0.025 \;{\rm GeV}^{-1}$
via PCAC arguments involving the AVV anomaly or alternatively, 
via the \lsm\ PVV quark loop \cite{twenty,twentyone}. 
The measured \cite{pdg} $\pi^0\gamma\gamma$ amplitude
is $0.025 \pm 0.001\;{\rm GeV}^{-1}$.} one extracts
\begin{equation}
|F_{K_L\gamma\gamma}|\cong \left|(\alpha/\pi
f_\pi)\frac{1}{(m_{K_L}^2-m_{\pi^0}^2)} \langle \pi^0|H_w^{\rm
pc}|K_L\rangle \right|=(3.49\pm 0.05)\times 10^{-9}\;{\rm
GeV}^{-1} \label{eq:tredici}
\end{equation}
from the lifetime $\tau_{K_L}=(5.17\pm 0.04)\times
10^{-8}$ sec and the branching ratio \cite{pdg} ${\cal
B}(K_L\to 2\gamma)=(5.86\pm 0.15)\times 10^{-4}$. Solving
(\ref{eq:tredici}) for the $K_L\to \pi^0$ weak transition, one
obtains 
\begin{equation}
|\langle \pi^0|H_w^{\rm pc}|K_L\rangle|=(3.20\pm 0.04)\times
10^{-8}\;{\rm GeV}^2. \label{eq:quattordici}
\end{equation}
If $\eta$ and $\eta^{\prime}$ pole contributions to $\langle 2 \gamma | H_w^{\rm pc}|K_L\rangle$ are taken into consideration, we
see from (A.12) of the Appendix that the central value of $|\langle \pi^0|H_w^{\rm pc}|K_L\rangle|$ 
in (14) will increase
by a multiplicative factor of $1/0.90$ to $3.56 \times 10^{-8}\;{\rm GeV}^2$.
In any case, the estimate (14) is quite close to its chiral-partner amplitude
(8), providing further phenomenological support for the relation
(10) anticipated from chiral symmetry.\footnote{A model independent (but non-chiral) estimate of $\langle
\pi^0|H_w|K_L\rangle=\sqrt{2}\langle \pi^0|H_w|K^0\rangle$, as obtained
from the meson self-energy type graphs of Fig. 4, yields
$|\langle \pi^0|H_w|K_L\rangle|=\left|\frac{G_F s_1
c_1}{\sqrt{2}}(m_D^2-m_K^2)\int^{\Lambda}\frac{d^4
p}{(2\pi)^4}\frac{p^2}{(p^2-m_D^2)(p^2-m_K^2)}\right|\approx
3.5\times 10^{-8}\;{\rm GeV}^2$ for the UV cutoff $\Lambda\approx m_D=1.87$ GeV \cite{tredici}.}

The analogue \si\ $L \sigma M$ pole contribution to $K_S\to 2\gamma$ in Fig. 3(b) yields
\begin{equation}
\langle 2\gamma|H_w^{\rm pv}|K_S\rangle=\langle
2\gamma|\sigma\rangle\frac{1}{m_{K_S}^2-m_{\sigma}^2+im_\sigma
\Gamma_\sigma}\langle\sigma|H_w^{\rm
pv}|K_S\rangle=F_{K_S\gamma\gamma}\epsilon^{\prime\mu}\epsilon^{\nu}(k_\mu
k^\prime_\nu-k\cdot k^{\prime}g_{\mu\nu}). \label{eq:quindici}
\end{equation}
Recall that $|1/(m_K^2 - m_\sigma^2 + i m_\sigma \Gamma_\sigma) | \simeq 1/m_\sigma^2$
for consistency of (6) with the chiral symmetry relation (10).
This constraint enables one to extract the strength
of the $\sigma \rightarrow \gamma\gamma$ amplitude within (15) from
measurable quantities:
\begin{equation}
<2\gamma | \sigma> \equiv F_{\sigma\gamma\gamma} \epsilon^{\prime\mu}
\epsilon^\nu \left( k_\mu k^\prime_\nu - k \cdot k^\prime
g_{\mu\nu}\right),
\label{sedici}
\end{equation}
\begin{equation}
|F_{\sigma\gamma\gamma} | \simeq | \frac{F_{K_S \gamma\gamma}  m_\sigma^2}{
< \pi^{0} | H_w^{pc} | K_L >}| = (5.5 \pm 1.6) \times 10^{-2}\;{\rm GeV}^{-
1}.
\label{eq:diciassette}
\end{equation}
The final numerical value in (17) is obtained for $m_\sigma = 600 \pm
50\;{\rm MeV}$ from (14) [or from (8) via (10)], $| < \pi^{\circ}|H_w^{pc}|K_L > | = 
3.56 \times 10^{-8} \; {\rm GeV}^2$, as discussed above, and the observed
$K_S \rightarrow \gamma\gamma$ amplitude $|F_{K_S \gamma\gamma}| =
(5.4 \pm 1.0) \times 10^{-9}\;{\rm GeV}^{-1}$ \cite{pdg}.
The result (17) leads to a $\sigma (600) \rightarrow \gamma\gamma$ width of
$m_\sigma^3|F_{\sigma \gamma\gamma}|^2 / 64 \pi = (3.2 \pm 2.2) {\rm keV}$, which
is compatible with a $3.8 \pm 1.5 {\rm keV}$ estimate \cite{twentytwo} deduced from
$\gamma\gamma \rightarrow \pi \pi$ data.

The estimate (17) is particularly useful for the weak decay $K_L
\rightarrow \pi^{0}\gamma\gamma$, which is anticipated (within a $L \sigma M$ context) 
to be dominated by the
$\sigma$(600) pole graph of Fig. 5.  We see from (17) that
\begin{equation}
|F_{\sigma\gamma\gamma}| = (2.20 \pm 0.64) \frac{\alpha}{\pi f_\pi}.
\label{eq:diciotto}
\end{equation}
The Figure 5 process has already been utilized \cite{quattordici} to predict
$\Gamma(K_L \rightarrow \pi^{0}\gamma\gamma) \approx 1.7 \times 10^{-
23}$ GeV with $F_{\sigma\gamma\gamma}$ assumed equal to $2\alpha / \pi f_\pi$.  
The corresponding prediction rescaled via (18) is
\begin{equation}
\Gamma(K_L \rightarrow \pi^{0}\gamma\gamma) = \left(2.1
_{-1.1}^{+1.3} \right) \times 10^{-23}\;{\rm GeV}.
\label{eq:diciannove}
\end{equation}
This prediction is insensitive to three-body
amplitudes (such as $K\to 3\pi$)
involving {\it two} amplitudes which partially subtract, due to
chiral symmetry, and is consistent with the measured rate
\cite{pdg,twentythree}
\begin{equation}
\Gamma (K_L\to \pi^0\gamma\gamma)=\frac{h}{2\pi\tau_{K_L}}(1.68\pm
0.07\pm 0.08)\times 10^{-6}=(2.15\pm 0.14)\times 10^{-23}\;{\rm
GeV}. \label{eq:venti}
\end{equation}

This rate, originally obtained from the \lsm\ Lagrangian
\cite{undici}, is useful for testing PCAC. The ratio of the rates
$K_S\stackrel{{\rm pv}}{\to} \gamma\gamma$ and $K_L\stackrel{{\rm
pc}}{\to} \pi^0\gamma\gamma$, as obtained from the $\sigma$-pole
graphs of Figs. 3b and 5, permits a cancellation of common
$\langle \sigma|H_w^{\rm pv}|K_S\rangle F_{\sigma \gamma\gamma}$
amplitude scales within each rate. To see this, first note that
the two amplitudes are related via a PCAC reduction of the $\pi^0$
state \cite{twentyfour}:
\begin{equation}
|\langle \gamma\gamma\pi^0|H_w^{\rm
pc}|K_L\rangle|=\frac{1}{f_\pi}|\langle
\gamma\gamma|[Q_5^3,H_w^{\rm
pc}]|K_L\rangle|=\frac{1}{2f_\pi}|\langle \gamma\gamma|H_w^{\rm
pv}|K_S\rangle|. \label{eq:newpr1}
\end{equation}
On the other hand the $\sigma$-pole graph of Fig. 5 provides a
direct determination of the $K_L\to\pi^0\gamma\gamma$ rate. Using
the three body phase space integral of ref. \cite{quattordici} for
$m_\sigma=640$ MeV, we find 
\begin{equation}
\Gamma(K_L\to\pi^0\gamma\gamma)=(1.7\times 10^{-4}\;{\rm
GeV}^4)|\langle\pi^0\sigma|H_w^{\rm pc}|K_L\rangle|^2
\frac{|F_{\sigma\gamma\gamma}|^2}{64 m_K^3 4\pi^3},
\label{eq:newpr2}
\end{equation}
whereas the $K_S\to\gamma\gamma$ amplitude in eq.
(\ref{eq:quindici})  corresponds to the rate
\begin{equation}
\Gamma(K_S\to\gamma\gamma)=\frac{m_K^3}{64\pi}|\langle\gamma\gamma|H_w^{\rm
pv}|K_S\rangle|^2. \label{eq:newpr3}
\end{equation}
Applying PCAC to $\langle\pi^0\sigma|H_w^{\rm pc}|K_L\rangle$ in
(\ref{eq:newpr2}) as in (\ref{eq:newpr1}), and again utilizing the
$\sigma$-pole graph of Fig. 3b in eq. (\ref{eq:newpr3}), we obtain
the following ratio of rates:
\begin{equation}
\frac{\Gamma(K_L\to\pi^0\gamma\gamma)}{\Gamma(K_S\to\gamma\gamma)}
\stackrel{{\rm
PCAC}}{=}\frac{(m_\sigma\Gamma_\sigma)^2}{m_K^6}\frac{(1.7\times
10^{-4} {\rm GeV}^4)}{(4\pi f_\pi)^2}=11.6\times 10^{-4},
\label{eq:newpr4}
\end{equation}
for $f_\pi\simeq 93 {\rm MeV}$, $m_\sigma=640$ MeV and, via eq.
(\ref{eq:undici}), $\Gamma_\sigma= 588$ MeV. Note the elimination
of the amplitude factor $\langle\sigma|H_w^{\rm pv}|K_S\rangle
F_{\sigma\gamma\gamma}$ within the ratio (\ref{eq:newpr4}). This
ratio is also seen to be compatible with data \cite{pdg},
\begin{equation}
\frac{\Gamma(K_L\to\pi^0\gamma\gamma)}{\Gamma(K_S\to\gamma\gamma)}\stackrel{{\rm
PDG}}{=}\frac{\frac{h}{2\pi\tau_L}(1.68\pm 0.10)\times
10^{-6}}{\frac{h}{2\pi\tau_S}(2.4\pm 0.9)\times 10^{-6}}=(12.1\pm
4.6)\times 10^{-4},
\end{equation}
given the measured lifetimes $\tau_L=(5.17\pm 0.04)\times 10^{-8}
$ sec, $\tau_S=(0.8935\pm 0.0008)\times 10^{-10}$ sec.
In effect the near-equivalence of (24) and (25), modulo the large error
in the latter result, provides confirmation for the PCAC
reduction employed in (21).

\begin{center}{\it Evidence for $\sigma$ in Three-Body Non-Leptonic Weak
Decays}
\end{center}

Consider the kaon three-body decay
$K^+\to\pi^+\sigma\to\pi^+\pi\pi$, for which the intermediate
$\sigma(600)$ is virtual. The
(model-dependent) dynamical graphs of Fig. 6, which partially cancel
due to chiral symmetry,
predict an analogue $K_L\to 3\pi^0$ decay rate $\Gamma\sim
3\times 10^{-18}\;{\rm GeV}$ \cite{tredici,quattordici} in rough agreement with data.
However, using (7) and PCAC consistency \cite{diciassette}, the $K^+\to 3\pi$
decay amplitude 
\begin{equation}
|\langle\pi^+\pi^-\pi^+|H_w|K^+\rangle|=\frac{1}{2 f_\pi} |\langle
\pi^+\pi^-|H_w|K_S\rangle|\approx 1.94\times 10^{-6},
\label{eq:ventuno}
\end{equation}
is in excellent agreement with the experimental amplitude \cite{pdg}
\begin{equation}
A_K=|\langle\pi^+\pi^-\pi^+|H_w|K^+\rangle|_{\rm exp}=(1.93\pm
0.01)\times 10^{-6} \label{eq:ventidue}
\end{equation}
obtained via the following three-body phase space integral \cite{diciassette}:
\begin{eqnarray}
\Gamma(K^+\to\pi^+\pi^-\pi^+)&=&\left(\frac{1}{8\pi
m}\right)^3|A_K|^2\int_{4\mu^2}^{(m-\mu)^2}ds \left[
\frac{(s-4\mu^2)(s-(m+\mu)^2)(s-(m-\mu)^2)}{s}\right]^{\frac{1}{2}}\nonumber\\
& &\nonumber\\
&=& I_K|A_K|^2 = (2.97 \pm 0.03) \times 10^{-18}\;{\rm GeV},
\label{eq:ventitre}
\end{eqnarray}
with $m,\mu$ being the kaon and pion masses respectively, and with
$I_K=0.798\times 10^{-6}\;{\rm GeV}$.

Note that we have used current-algebra/PCAC consistency,
as evident from the agreement between
(\ref{eq:ventuno}) and (\ref{eq:ventidue}), to infer that
the $\sigma$ and $\pi$ mesons in Fig. 6 are chiral partners. In
effect, we have factored $|A_K|^2$ from the integral in eq.
(\ref{eq:ventitre}), treating $|A_K|$ as independent of $s$
because the underlying $\sigma$-pole graph in Fig. 6 is
virtual in the transition $K^+\to \pi^+\sigma\to\pi^+\pi\pi$.

Evidence for a {\it non-virtual} $\sigma$ may be extracted by
relating the non-resonant  fraction of the $D^+\to\pi^+\pi^+\pi^-$
three body decay to the known two body
$|\langle\pi^+\pi^-|H_w|D^0\rangle|$ matrix element via analyses
which either {\it do} or {\it do not} include an
appreciable $\pi^+\sigma$-resonant contribution. The former case
corresponds to a fit obtained by the E791 Collaboration
\cite{sette} with an apparent $\sigma$ mass of $478^{+24}_{-23}\pm
17\;{\rm MeV }$ and width of $324^{+42}_{-40}\pm 21\;{\rm MeV}$.
The latter case is implicit within the PDG  estimate for the non-resonant (NR) 
contribution to the
$D^+\to\pi^+\pi^-\pi^+$ rate \cite{pdg}:
\begin{equation}
\Gamma^{\rm PDG}_{\rm
(NR)}(D^+\to\pi^+\pi^-\pi^+)=\frac{h}{2\pi\tau_{D^+}} (2.2\pm
0.4)\times 10^{-3}=(1.37\pm 0.25)\times 10^{-15}\;{\rm GeV}.
\label{eq:ventiquattro}
\end{equation}
This corresponds to $(2.2\pm 0.4)/(3.6\pm 0.4)=(61\pm 18)\%$ of the
total $D^+\to \pi^+\pi^+\pi^-$ decay rate. Since the
$\rho^0\pi^+$ resonant-channel branching fraction is $(1.05\pm
0.31)/(3.6\pm 0.4)=(29\pm 12)\%$, the PDG
non-resonant contribution (\ref{eq:ventiquattro}) is essentially
the difference between the full $D^+\to\pi^+\pi^+\pi^-$ rate and
the $D^+\to\pi^+\rho^0\to\pi^+\pi^+\pi^-$ resonant sub-rate; the
$D^+\to\pi^+\sigma\to\pi^+\pi^+\pi^-$ resonant sub-rate is at best
assumed by the PDG to be a secondary contribution within the
$10\%$ remaining for resonance sub-rates other than $\rho$.
Assuming the rate (\ref{eq:ventiquattro}) is truly due to
non-resonant virtual intermediate states (which presumably are
independent of the squared energy variable $s$), we find that
\begin{equation}
\Gamma^{\rm PDG}_{\rm
(NR)}=J_{D^+}|\langle\pi^+\pi^+\pi^-|H_w|D^+\rangle_{\rm NR}|^2,
\label{eq:venticinque}
\end{equation}
where the constant squared amplitude has been factored out from
the three body phase space integral (analogous to
(\ref{eq:ventitre})) whose numerical value \cite{diciassette} is
$J_{D^+}=48.8\times 10^{-6}\;{\rm GeV}$. Substitution of eq.
(\ref{eq:ventiquattro}) into (\ref{eq:venticinque}) then predicts
the non-resonant amplitude
\begin{equation}
|\langle\pi^+\pi^+\pi^-|H_w|D^+\rangle_{\rm NR}|=(5.3\pm
0.5)\times 10^{-6}, \label{eq:ventisei}
\end{equation}
which leads, via PCAC (as in
(\ref{eq:newpr4}),(\ref{eq:ventuno})), to the prediction
\begin{equation}
|\langle\pi^+\pi^-|H_w|D^0\rangle|\approx\sqrt{2}
f_\pi|\langle\pi^+\pi^+\pi^-|H_w|D^+\rangle_{\rm
NR}|=(7.0\pm 0.6)\times 10^{-7}\;{\rm GeV }. \label{eq:ventisette}
\end{equation}
Comparison of this prediction to the empirical value for this
matrix element
\begin{equation}
|\langle\pi^+\pi^-|H_w|D^0\rangle|=m_{D^0}\sqrt{\frac{8\pi\Gamma(D^0\to\pi^+\pi^-)}{q}}=(4.8\pm
0.2)\times 10^{-7}\;{\rm GeV}, \label{eq:ventiotto}
\end{equation}
(where $q=922$ MeV and $\Gamma (D^0\to\pi^+\pi^-)\approx 24\times
10^{-16}$ GeV \cite{pdg}) suggests that the PDG value
(\ref{eq:ventiquattro}) for the non-resonant $D^+\to\pi^+\pi^+\pi^-$
rate may be too large. Such would be the case if it failed to take into account a
$D^+\to\pi^+\sigma\to\pi^+\pi^+\pi^-$ resonant contribution.

As noted above, such a contribution appears to be evident in the
recent E791 Collaboration measurements of the
$D^+\to\pi^+\pi^+\pi^-$ rate, leading to a fit in which $46.3\%$
of the rate occurs via $D^+\to\pi^+\sigma$, and that $33.6\%$ of
the rate occurs via $D^+\to\pi^+\rho^0$. This latter branching
fraction is consistent with the $(29\pm 12)\%$ Particle
Data Group branching fraction for the $\pi^+\rho^0$ resonant
channel. Hence, the net effect of the new E791 data is essentially
to reduce the non-resonant fraction of $D^+\to\pi^+\pi^+\pi^-$
from its PDG value $(61\pm 18)\%$ to no more than $20\%$
($=1-(33.6\%)_{\pi^+\rho}-(46.3\%)_{\pi^+\sigma}$). Using this
upper bound, the non-resonant rate is sharply reduced from
(\ref{eq:ventiquattro}) to
\begin{equation}
\Gamma_{({\rm NR}\;{\rm
no}\;\sigma)}=0.20\times\frac{h}{2\pi\tau_{D^+}}{\cal
B}(D^+\to\pi^+\pi^+\pi^-)=(4.48\pm 0.5)\times 10^{-16}\;{\rm GeV},
\label{eq:ventinove}
\end{equation}
as obtained from the PDG branching ratio ${\cal
B}(D^+\to\pi^+\pi^+\pi^-)=(3.6\pm 0.4)\times 10^{-3}$. Eq.
(\ref{eq:ventinove}) leads via PCAC to a prediction for the
$D^0\to\pi^+\pi^-$ matrix element that appears to be consistent with the
experimental value (\ref{eq:ventiotto}),
\begin{equation}
\langle\pi^+\pi^-|H_w|D^0\rangle\approx\sqrt{2}f_\pi\sqrt{\frac{
\Gamma_{({\rm NR}\;{\rm no}\;\sigma)}}{J_{D^+}}}=(4.0\pm
0.2)\times 10^{-7}\;{\rm GeV}. \label{eq:trenta}
\end{equation}
The reasonable agreement between eqs. (\ref{eq:ventiotto}) and
(\ref{eq:trenta}) is evidence for the production of a {\it physical} isoscalar  $\sigma(600)$
meson within the $D^+\to \pi^+\pi^+\pi^-$ weak
decay.

This agreement is subject to two experimental caveats: 1)
the non-resonant rate (\ref{eq:ventinove}) may in fact have other
resonant processes buried within it besides the dominant
$D^+\to\pi^+\sigma$ and $D^+\to\pi^+\rho^0$ resonant channels,
\footnote{The E791 collaboration central value estimate for
$\Gamma_{({\rm NR})}(D^+\to \pi^+\pi^+\pi^-)$ is only $7.8\%$, due
to an additional contribution from resonances other than $\rho$
and $\sigma$. Using this estimate, the eq. (\ref{eq:trenta}) matrix element is reduced to
$2.5\times 10^{-7} \; GeV$.} and 2) the PDG branching fraction quoted
above for the full $D^+\to\pi^+\pi^+\pi^-$ rate may in fact be too
large. This second possibility is suggested by the E791
Collaboration's new measurements of $\Gamma
(D^+\to\pi^+\pi^+\pi^-)/\Gamma (D^+\to\pi^+\pi^+ K^-)=0.0311\pm
0.018^{+0.0016}_{-0.0026}$. The $D^+\to\pi^+\pi^+ K^-$ branching
fraction $(9.0\pm 0.6)\%$ \cite{pdg} suggests a lowering of ${\cal
B}(D^+\to\pi^+\pi^+\pi^-)$ to $(2.96^{+0.48}_{-0.57})\times
10^{-3}$, a reduction of $\sim 20\%$ from the PDG value $(3.6\pm
0.4)\times 10^{-3}$. Such a $20\%$ reduction in the
$D^+\to\pi^+\pi^+\pi^-$ branching fraction would reduce the
prediction of (\ref{eq:trenta}) by $10\%$.

\acknowledgements

ADP and NAT acknowledge support  from EU-TMR programme, contract CT98-0169 and
ADP is grateful to M.L. Mangano for his kind hospitality at CERN.
VE is grateful for research support from the Natural Sciences and Engineering
Research Council of Canada and for hospitality from The University of Arizona.
MDS acknowledges prior work with A. Bramon, R. E. Karlsen and S.R. Choudhury.

\section*{Appendix: $\eta$- and $\eta^{\prime}$-Pole Contributions to 
$K_L \rightarrow \gamma\gamma$}

If the $K_L \rightarrow \gamma\gamma$ amplitude of Fig. 3a is augmented
by contributions from $\eta$ and $\eta^{\prime}$ poles, the $\pi^{\circ}$,
$\eta$, and $\eta^{\prime}$  contributions to the amplitude are respectively
given by

\renewcommand {\theequation}{A.\arabic{equation}}
\setcounter{equation}{0}

\begin{equation}
M_{\pi^{\circ}} = <2\gamma | \pi^{\circ}> \frac{1}{m_{K_L}^2 -
m_{\pi^{\circ}}^2} < \pi^{\circ} | H_w^{pc} | K_L>,
\label{eq:A.1}
\end{equation}

\begin{equation}
M_\eta = <2\gamma | \eta> \frac{1}{m_{K_L}^2 -
m_\eta^2} < \eta | H_w^{pc} | K_L>,
\label{eq:A.2}
\end{equation}
\begin{equation}
M_{\eta^{\prime}} = <2\gamma | \eta^{\prime} > \frac{1}{m_{K_L}^2 -
m_{\eta^{\prime}}^2} < \eta^{\prime} | H_w^{pc} | K_L>.
\label{eq:A.3}
\end{equation}
To find the relative contributions of these matrix elements, we first
note that 
\begin{equation}
<\eta | H_w | K_L> = cos \theta_p < \eta_8 | H_w | K_L > - sin \theta_p
< \eta_0 | H_w | K_L >,
\label{eq:A.4}
\end{equation}
\begin{equation}
<\eta^{\prime} | H_w | K_L> = sin \theta_p < \eta_8 | H_w | K_L > + cos \theta_p
< \eta_0 | H_w | K_L >,
\label{eq:A.5}
\end{equation}
where the pseudoscalar mixing angle $\theta_p = -12.9^{\circ}$
\cite{twentyfive,twentysix}.  If the relative sizes of transitions
from $K_L$ to nonstrange pseudoscalar-nonet states is 
scaled to the $U(3)$ structure constants [{\it i.e.},
$\left( <\pi^{\circ} | H_w^{pc} | K_L > : \right.$ $ < \eta_8 | H_w^{pc} | K_L > :$
$\left. <\eta_0 | H_w ^{pc} | K_L > \right)\; = \; \left( d_{366} \; : \;
d_{866} \; : \; d_{066} \right) = \left( - \frac{1}{2} \; : \; -\frac{1}{2 \sqrt{3}} \;
: \; \sqrt{\frac{2}{3}} \right)$], we then find that
\begin{equation}
< \eta | H_w | K_L > = 0.198 < \pi^{\circ} | H_w | K_L > ,
\label{eq:A.6}
\end{equation}
\begin{equation}
< \eta^{\prime} | H_w | K_L > = -1.72 < \pi^{\circ} | H_w | K_L > .
\label{eq:A.7}
\end{equation}
Using the matrix elements \cite{twentysix}
\begin{eqnarray}
< 2 \gamma | \pi^{\circ} > = 0.0250 \; GeV^{-1}, \nonumber\\
< 2 \gamma | \eta > = 0.0255 \; GeV^{-1}, \nonumber\\
< 2 \gamma | \eta^{\prime} > = 0.0335 \; GeV^{-1}
\label{eq:A.8}
\end{eqnarray}
[Levi-Civita covariants have been factored out of (A.8)], we then find
from (A.1-3) that the matrix elements for $\pi^{\circ}$, $\eta$, and
$\eta^{\prime}$ pole contributions to $K_L \rightarrow 2 \gamma $ are
respectively given by
\begin{equation}
M_{\pi^{\circ}} = (0.109 \; GeV^{-3}) < \pi^{\circ} | H_w^{pc} | K_L >,
\label{eq:A.9}
\end{equation}
\begin{equation}
M_\eta = (-0.0975 \; GeV^{-3}) < \pi^{\circ} | H_w^{pc} | K_L >,
\label{eq:A.10}
\end{equation}
\begin{equation}
M_{\eta^{\prime}} = (+0.0861 \; GeV^{-3}) < \pi^{\circ} | H_w^{pc} | K_L >.
\label{eq:A.11}
\end{equation}
Consequently, there is a near cancellation of $\eta$ and $\eta^{\prime}$ pole
contributions in the matrix-element sum:
\begin{equation}
M_{\pi^{\circ}} + M_\eta + M_{\eta^{\prime}} = (0.0976 \; GeV^{-3}) <
\pi^{\circ} | H_w^{pc} | K_L > = (0.90) M_{\pi^{\circ}} .
\label{eq:A.12}
\end{equation}

\newpage
\begin{figure}[t!]
\begin{center}
\epsfig{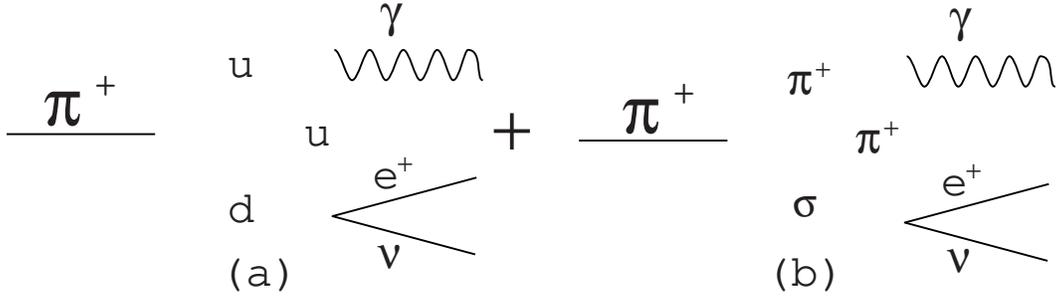}
\caption{\label{fig:fig1}
\footnotesize
          \lsm\ quark (a) and meson (b) loops for $\pi^+\to e^+\nu\gamma$ decay. }
\end{center}
\end{figure}
\begin{figure}[t!]
\begin{center}
\epsfig{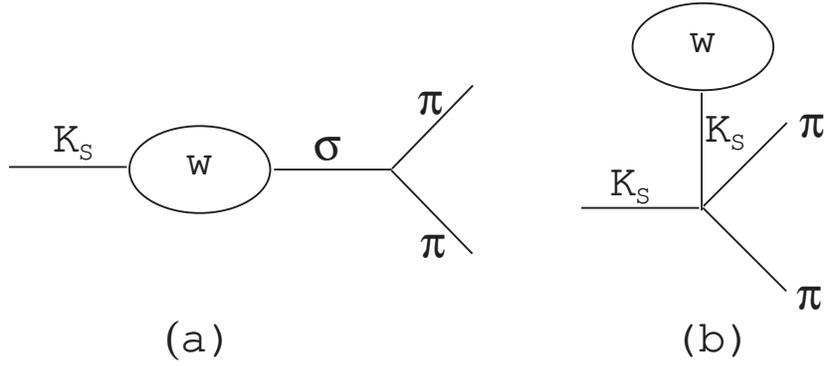}
\caption{\label{fig:fig2}
\footnotesize
          $K_S\to \pi\pi$ $\Delta I=\frac{1}{2}$ tree graphs in the $s$-channel (a) or
          in the $t$-channel (b). }
\end{center}
\end{figure}
\begin{figure}[t!]
\begin{center}
\epsfig{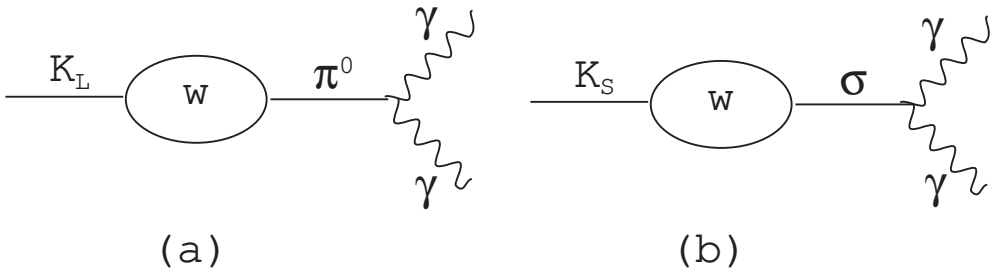}
\caption{\label{fig:fig3}
\footnotesize
         $K_L\to 2\gamma$ decay dominated by a $\pi^0$ pole (a),
         $K_S\to 2\gamma$ decay dominated by a $\sigma$ pole (b). }
\end{center}
\end{figure}
\begin{figure}[t!]
\begin{center}
\epsfig{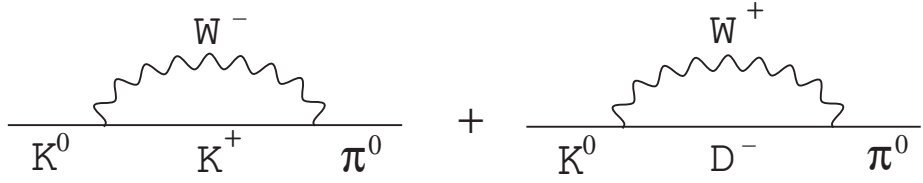}
\caption{\label{fig:fig4}
\footnotesize
          Self energy-type $W$ graphs for the $K^0\to \pi^0$ weak transition. }
\end{center}
\end{figure}
\begin{figure}[t!]
\begin{center}
\epsfig{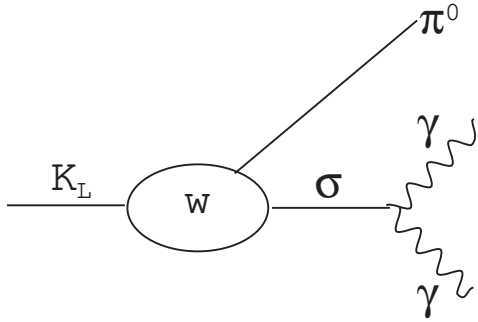}
\caption{\label{fig:fig6}
\footnotesize
          $K_L\to \pi^0\gamma\gamma$ weak decay via $\sigma$ pole. }
\end{center}
\end{figure}
\begin{figure}[t!]
\begin{center}
\epsfig{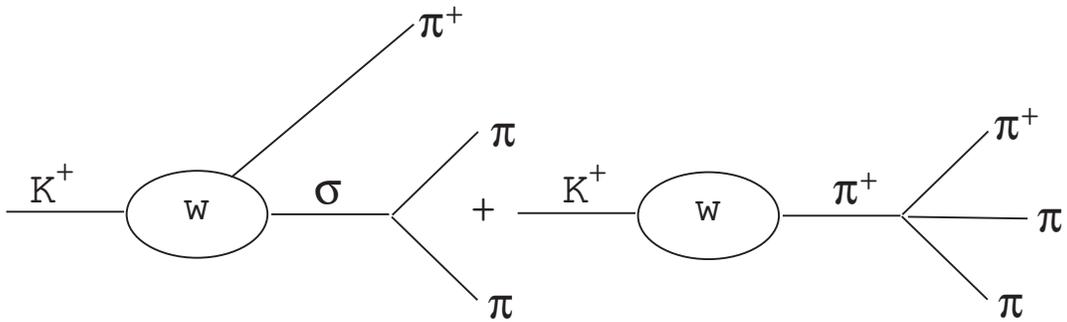}
\caption{\label{fig:fig7}
\footnotesize
          $K^+\to\pi^+\pi\pi$ decay via chirally subtracted $\sigma$
          and $\pi^+$ pole graphs. }
\end{center}
\end{figure}


\begin{thebibliography}{10}

\bibitem{pdg} C.Caso et al. (The Particle Data Group) Eur.Phys. J {\bf C3}, 1 (1998).

\bibitem{due} N.A. T\"ornqvist and M. Roos, Phys. Rev. Lett. {\bf 76}, 1575
(1996); N.A. T\"ornqvist Zeit. Physik {\bf C68}, 647 (1995); S.
Ishida et al. Prog. Theor. Phys. {\bf 95}, 745 (1995); R. Kaminski
et al. Phys. Rev. {\bf D50}, 3145 (1994); V. Elias and M. D. Scadron, Phys. Rev.
Lett.{\bf 53}, 1129 (1984); R. Delbourgo and M.D. Scadron,
Phys. Rev. Lett. {\bf 48}, 379 (1982) and Mod. Phys. Lett. {\bf A10},
251 (1995).

\bibitem{tre} E. van Beveren et al. Zeit. Phys. {\bf C30}, 616 (1986);
J.E. Augustin et al. Nucl. Phys. {\bf B320}, 1 (1989);
J.A. Oller et al. Nucl. Phys. {\bf A620}, 438 (1997);
R. Kaminski et al. Z. Phys. {\bf C74}, 79 (1997) and
Eur. Phys. J. {\bf C9}, 141 (1999);
M.P. Locher et al. Eur. Phys. J. {\bf C4}, 317 (1998);
D. Black et al. Phys. Rev. {\bf D58}, 054012 (1998);
R. Delbourgo and M. Scadron, Int. J. Mod. Phys. {\bf A13}, 657(1998);
K. Igi and K. Hikasa, Phys. Rev. {\bf D59}, 034005 (1999);
J.A. Oller and E. Oset, Phys. Rev. {\bf D59}, 074001 (1999) and
Phys. Rev. {\bf D60},  074023 (1999);
M. Scadron, Eur. Phys. J. {\bf C6}, 141 (1999);
N.A. T\"ornqvist Eur. Phys. J. {\bf C11}, 359 (1999);
T. Hannah, Phys. Rev. {\bf D60}, 017502 (1999);
CLEO Collaboration, Phys. Rev. {\bf D61} (2000) 012002.

\bibitem{quattro}M. Harada, F. Sannino and J. Schechter, Phys. Rev. {\bf D54}, 1991 (1996).

\bibitem{cinq}F. Sannino and J. Schechter, Phys. Rev. {\bf D52}, 96 (1995).

\bibitem{sei} N.Isgur and J. Speth, Phys. Rev. Lett. {\bf 77},
2332 (1996);
N.A. T\"ornqvist and M. Roos, ibid. {\bf 77}, 2333 (1996).

\bibitem{sette} E. M. Aitala et al. (E791 collaboration) {\it Experimental
evidence for a light and broad scalar resonance in
$D^+\to\pi^-\pi^+\pi^+$ decay}, Phys. Rev. Lett. {\bf 86}, 770 (2001).


\bibitem{otto} See e.g. N.Paver and M.D. Scadron, Nuovo Cimento {\bf A78}, 159 (1983);
Ll- Ametller, C. Ayala and  A. Bramon, Phys.Rev {\bf D29}, 916
(1984).

\bibitem{nove} A. Bramon and M.D. Scadron, Europhys. Lett. {\bf
19}, 663 (1992); A. Bramon, R.E. Karlsen and M.D. Scadron, Mod.
Phys. Lett. {\bf A8}, 97 (1993); also see P. Pascual and R. Tarrach,
Nucl. Phys. {\bf B146}, 509 (1978) and S. B. Gerasimov, Sov. J. Nucl. Phys.
 {\bf 29}, 259 (1979).  For a current-algebra estimate of $F_A / F_V = 0.6$, see
T. Das, V. S. Mathur and S. Okubo, Phys. Rev. Lett. {\bf 19}, 859 (1967).

\bibitem{dieci} T. Marozuni, C.S. Lim and A.I. Sanda, Phys. Rev.
Lett. {\bf 65}, 404 (1990).

\bibitem{undici} M. Gell-Mann and M. Levy, Nuovo Cimento {\bf 16},
705 (1960); also see V. de Alfaro, S. Fubini, G. Furlan and C.
Rossetti in {\it Currents in Hadron Physics} (North Holland,
1973), chap. 5.

\bibitem{dodici} S. Weinberg, Phys. Rev. Lett. {\bf 65}, 1177
(1990); M.D. Scadron, Mod. Phys. Lett. {\bf A7}, 497 (1992); P. Ko
and S. Rudaz, Phys. Rev. {\bf D50}, 6877 (1994).

\bibitem{tredici} R.E. Karlsen and M. D. Scadron, Mod. Phys. Lett.
{\bf A6}, 543 (1991).

\bibitem{quattordici} R.E. Karlsen and M.D. Scadron, Nuovo Cimento {\bf
A106}, 113 (1993).

\bibitem{quindici} R.E. Karlsen and M.D. Scadron, Phys. Rev.
{\bf D44}, 2192 (1991).

\bibitem{sedici} G. Eilam and M.D. Scadron, Phys.
Rev. {\bf D31}, 2263 (1985); S.R. Choudhury and M. D. Scadron,
Nuovo Cimento {\bf A108}, 289 (1995).

\bibitem{diciassette} R.E. Karlsen and M.D. Scadron, Phys. Rev. {\bf
D45}, 4108 (1992); {\bf D45}, 4113 (1992).

\bibitem{diciotto} S. Weinberg, Phys. Rev. Lett {\bf 17}, 616 (1966);
H. Osborn, Nucl. Phys. {\bf B15}, 501 (1970).

\bibitem{nineteen} M.D. Scadron, Repts. on Prog. Phys. {\bf 44}, 213
(1981).

\bibitem{twenty}J. Steinberger, Phys. Rev. {\bf 76}, 1180 (1949).

\bibitem{twentyone}A. S. Deakin, V. Elias and M.D. Scadron, Mod. Phys.
Lett. {\bf A9}, 955 (1994).

\bibitem{twentytwo}M. Boglione and M. R. Pennington, Eur. Phys. J. {\bf C9}, 11 (1999).

\bibitem{twentythree}{\it KTeV Collab}. [A. Alavi-Harati {\it et al}.], Phys.
Rev. Lett. {\bf 83}, 917 (1999).

\bibitem{twentyfour}A. Della Selva, A. De Rujula, and M. Mateev, Phys. Lett. {\bf B24}, 468 (1967).

\bibitem{twentyfive}H. F. Jones and M. D. Scadron, Nucl. Phys. {\bf
B155}, 409 (1979); M. D. Scadron, Phys. Rev. {\bf D29}, 2076 (1984);
for a review, see T. Feldmann, Int. J. Mod. Phys. {\bf A15}, 159 (2000).

\bibitem{twentysix}R. Delbourgo, Dongsheng Liu and M. D. Scadron, Int.
J. Mod. Phys. {\bf A14}, 4331 (1999).

\end{thebibliography}
\end{document}